\documentclass[%
 reprint,
 amsmath,amssymb,
 aps,
prl,longbibliography,
]{revtex4-2}

\usepackage{graphicx}
\usepackage{dcolumn}
\usepackage{bm}
\usepackage{hyperref}
\hypersetup{colorlinks=true,breaklinks,linkcolor=blue,urlcolor=blue,citecolor=blue}
\usepackage[mathlines]{lineno}
\usepackage{algorithm}
\usepackage[noend]{algpseudocode}
\usepackage{xcolor}
\usepackage{pifont}
\usepackage{tikz}
\usetikzlibrary{matrix}
\usepackage[normalem]{ulem}
\usepackage{verbatim}
\usepackage{soul}
\usepackage{mathtools}
\usepackage{ulem}
\usepackage[utf8]{inputenc}\DeclareUnicodeCharacter{2212}{-}

\begin{document}
\preprint{arXiv}

\title{Magnetism and Metallicity in Moir{\'e} Transition Metal Dichalcogenides}

\author{Patrick Tscheppe$^{1,2}$}
\author{Jiawei Zang$^{3}$}
\author{Marcel Klett$^{1}$}
\author{Seher Karakuzu$^{4}$}
\author{Armelle Celarier$^{5}$}
\author{Zhengqian Cheng$^{6}$}
\author{Chris A. Marianetti$^{6}$}
\author{Thomas A. Maier$^{7}$}
\author{Michel Ferrero$^{5,8}$}
\author{Andrew J. Millis$^{3,4}$}
\author{Thomas Sch{\"a}fer$^{1}$}
\email{t.schaefer@fkf.mpg.de}
 
\affiliation{
$^1$Max-Planck-Institut für Festkörperforschung, Heisenbergstraße 1, 70569 Stuttgart, Germany}
\affiliation{$^2$Institut für Theoretische Physik and Center for Quantum Science, Universität T{\"u}bingen, Auf der Morgenstelle 14, 72076 T{\"u}bingen, Germany}
\affiliation{
$^3$Department of Physics, Columbia University, 538 West 120th Street, New York, New York 10027, USA}
\affiliation{
$^4$Center for Computational Quantum Physics, Flatiron Institute, 162 5th Avenue, New York, New York 10010, USA}
\affiliation{$^5$CPHT, CNRS, {\'E}cole Polytechnique, Institut Polytechnique de Paris, Route de Saclay, 91128 Palaiseau, France}
\affiliation{
$^6$Department of Applied Physics and Applied Mathematics, Columbia University, New York, New York 10027, USA}
\affiliation{$^7$Computational Sciences and Engineering Division, Oak Ridge National Laboratory, Oak Ridge, Tennessee 37831-6164, USA}
\affiliation{$^8$Coll\`ege de France, 11 place Marcelin Berthelot, 75005 Paris, France \\}

\date{\today}

\begin{abstract}
The ability to control the properties of twisted bilayer transition metal dichalcogenides in situ makes them an ideal platform for investigating the  interplay of strong correlations and geometric frustration. Of particular interest are the low energy scales, which make it possible to experimentally access both temperature and magnetic fields that are of the order of the bandwidth or the correlation scale. In this manuscript we analyze the moir{\'e} Hubbard model, believed to describe the low energy physics of an important subclass of the twisted bilayer compounds. We establish its magnetic and the metal-insulator phase diagram for the full range of magnetic fields up to the fully spin polarized state. We find a rich phase diagram including fully and partially polarized insulating and metallic phases of which we determine the interplay of magnetic order, Zeeman-field, and metallicity, and make connection to recent experiments.
\end{abstract}

\maketitle

{\it Introduction.}
The correlation-driven Mott metal-insulator transition --in other words, under what circumstances can electrons move through a material-- is one of the central issues in modern day condensed matter physics. Developments over the past five years in moir{\'e} transition metal dichalcogenides, including the observation of a continuous Mott transition \cite{Li2021} and quantum criticality \cite{Ghiotto2021}, have opened a new experimental frontier in this area \cite{Tang2020,Wang2020,Kennes2021,Wu2018}. Moir{\'e} materials consist of two or more atomically thin layers, perhaps with slightly different lattice constants, stacked at a small twist angle. The lattice mismatch and twist angle, combined with a weak but nonzero interlayer tunnelling, produce experimental platforms whose low energy physics is described by a few-band model with a very large unit cell and therefore very low bandwidth and interaction scales, which moreover are tunable by twist angle, pressure, and the choice of materials in which the moir{\'e} system is embedded \cite{Kennes2021,Tang2020}. One particularly widely studied class of moir{\'e} materials are bilayers comprised of transition metal dichalcogenide materials such as WSe$_2$ and MoTe$_2$ which in appropriate circumstances realize the {\sl moir{\'e} Hubbard model}: a two-dimensional triangular lattice hosting a single band of electrons correlated by an  interaction that to a good approximation may be taken to be site-local. Importantly, the magnitude and form of the interaction and the electronic band structure can be varied over wide ranges {\sl in situ} by changing gate potentials and twist angles \cite{Wu2018,Wu2019,Pan2020,Wang2020,Kennes2021} while all electronic scales are small enough that temperatures and magnetic fields spanning the whole range from very low to higher than the effective bandwidth are experimentally accessible.

While the metal-insulator transition in two dimensional Hubbard models has been studied, both in general \cite{Georges1996,Parcollet2004,Moukouri2001,Schaefer2015,Schaefer2016b,Wietek2021,Klett2020,Downey2022} and in connection to moir{\'e} systems \cite{Morales2021,Zang2022}, the effect of a magnetic field seems (apart from one notable exception \cite{Laloux94}) not to have been investigated, perhaps in part because for most conventional materials the experimentally accessible fields are a tiny fraction of the bandwidth so that linear response theory suffices. Motivated by the wide range of field strengths experimentally accessible in moir{\'e} systems,  in this paper we use state of the art single-site and cluster dynamical mean-field methods to study the metal-insulator phase diagram of the moir{\'e} Hubbard model over the full magnetic field range, assuming that the primary coupling is the Zeeman-coupling to the electronic spin. Orbital effects~\cite{Acheche2017,Markov2019,Vucicevic2021,Vucicevic2021b} will be considered in a forthcoming paper.  We reveal full and partially polarized insulating and metallic phases, as well as canted antiferromagnetically ordered phases.

\begin{figure}
		\centering
		\includegraphics{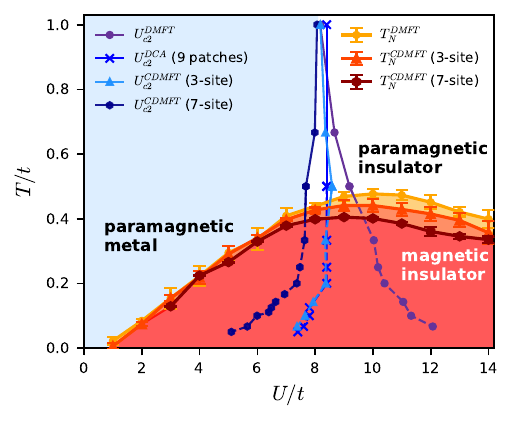}
		\caption{Phase diagram of the moir{\'e} Hubbard model (Eq.~\ref{eq:mhm}) for $\phi=\pi/6$ at half-filling and zero external magnetic field $B\!=\!0$ calculated by the dynamical mean-field methods indicated in the legends. Solid reddish lines denote magnetic phase transition lines from a paramagnetic to a 120$^\circ$ ordered antiferromagnetic state, solid blueish lines denote a crossover from a metallic to an insulating region. Dashed blueish lines mark a metal-insulator crossover when the calculation is restricted to a non-magnetic (metastable) solution.}
		\label{fig:phasediagram}
\end{figure}

{\it Model and methods. \label{sec:model}}
We study the  moir{\'e} Hubbard model (MHM), a modification of the well-known Hubbard model \cite{Hubbard1963, Hubbard1964, Gutzwiller1963, Kanamori1963}, the fruit fly of electronic correlations \cite{Qin2022,Arovas2022}. The Hamiltonian is
\begin{eqnarray}
    H&=&\!-\!\sum\limits_{\left<ij\right>,\sigma={\uparrow\downarrow}}c_{i,\sigma}^{\dagger}t_{\sigma}^{ij}c_{j,\sigma}+U\sum\limits_{i}n_{i,\uparrow}n_{i,\downarrow}\nonumber\\&&- g\mu_B B \sum_{i,\alpha,\beta} c^{\dagger}_{i,\alpha} \sigma_z^{\alpha, \beta} c_{i,\beta}.
    \label{eq:mhm}
\end{eqnarray}  Here, $i,j$ represent nearest-neighbor sites on a two-dimensional triangular lattice, $U$ is the (purely local) Coulomb interaction, and $t_{\sigma}^{ij}\!=\!\left|t\right|e^{i\sigma\phi_{ij}}$ is a spin-dependent hopping parameter \cite{Zang2021}, which can be parameterized by a complex phase $\phi$ arising from the strong spin-orbit coupling of the constituent layers and a magnitude $t$. $g$ is the gyromagnetic factor of an electron, $\mu_\text{B}$ is the Bohr magneton, and $B$ an externally applied field in $z$-direction. The structure of the model is such that at $\phi=\pi/6$ at half-filling the model has a particle-hole symmetry, a nested Fermi surface, and a third-order van Hove point, implying that at  $T\!=\!B\!=\!0$ the system is a $120^{\circ}-$antiferromagnetic insulator at even infinitesimal coupling; while for $\phi\neq \pi/6$ the model at $T=0$ is a paramagnetic metal at small interaction strengths, with a first order magnetic and metal-insulator transition as the relative interaction $U/t$ is increased above a critical value \cite{Zang2021}. Both $t$ and $\phi$ may be experimentally tuned {\sl in situ} by the application of appropriate gate voltages. In this work we analyze the half-filled situation $\left<n_\uparrow\right>\!+\!\left<n_\downarrow\right>\!=\!1$, considering both paramagnetic and $120^\circ$ magnetically ordered states. 

We investigate this model by means of the dynamical mean-field theory (DMFT \cite{Metzner1989,Georges1992,Georges1996}),  in its single site and cluster forms. We employ two flavors of  cluster DMFT: the cellular DMFT (CDMFT \cite{Maier2005} with center-focused post-processing \cite{Klett2020}) and the dynamical cluster approximation (DCA \cite{Maier2005}). We use cluster sizes $N_c\!\in\!\left\{1,3,7,9\right\}$. For our calculations at non-zero temperatures we use continuous-time quantum Monte Carlo in its interaction expansion (CT-INT), using the TRIQS package \cite{TRIQS}, to solve the dynamical mean-field equations \cite{Rubtsov2005,Gull2011}. These methods provide results only above a certain low temperature limit, which is typically low enough that a reliable extrapolation to the $T=0$ physics is possible. For some of our calculations we employ the recently developed variational discrete action theory (VDAT) \cite{Cheng2021,Cheng21,Cheng2022} which provides an extremely computationally efficient estimate of ground state properties of the single-site model. Details of the solvers,  cluster geometries and implementations are given in the Supplemental Material \cite{Supplemental}.

\begin{figure*}[t!]
		\centering
		\includegraphics{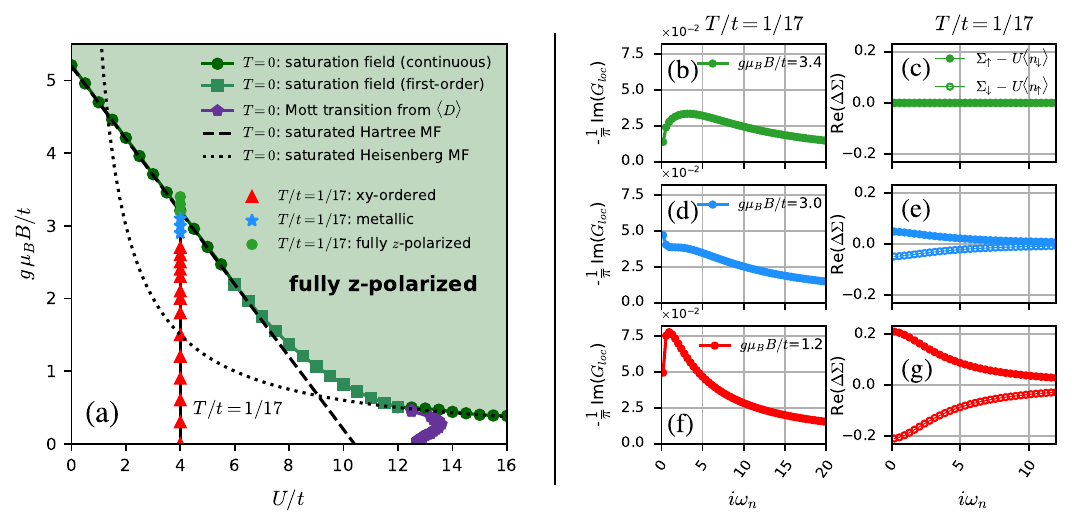}
		\caption{(a) Zero temperature phase diagram for the half-filled perfectly nested MHM ($\phi=\pi/6$), calculated in the interaction strength ($U$)-magnetic field ($B$) plane in the single-site dynamical mean-field approximation at $T=0$ using the VDAT method in the paramagnetic phase, i.e., without permitting spontaneous ordering. A phase boundary separates a large $B$, fully spin-polarized trivially insulating phase (green shaded region) from a small $B$ partially spin-polarized phase. For $U/t\lesssim 6$ and $U/t\gtrsim 12$ the transition to the fully polarized phase is second order (green circles) while for ($6\lesssim U/t\lesssim 12$)  the transition is first order (green squares). The dashed line shows the Hartree-Fock approximation to the full-polarization transition line. The dotted line shows the mean-field approximation to the full polarization transition of a nearest-neighbor Heisenberg model with $J\propto t^2/U$. Also shown is the critical coupling  $U_c$ of the Mott-Hubbard metal-insulator transition, which is seen to be reentrant as a function of field (purple pentagons). The vertical line at $U/t=4$ indicates $B$ fields at which a dynamical mean-field calculation at a temperature $T/t=1/17\approx 0.06$ leads to an antiferromagnetic insulator (red triangles), paramagnetic metal (blue stars) or trivial fully spin-polarized insulator (green circles). (b)-(g) Single-site dynamical mean-field results for the Matsubara frequency dependence of the imaginary part of the Green function (averaged over spin) and the two spin components of the dynamical self energy $\Delta \Sigma=\Sigma(\omega)-\Sigma(\omega\!\rightarrow\!\infty)$ obtained at the nonzero temperature $T/t=1/17$ for several  magnetic fields along the vertical $U/t=4$ line of (a).}
		\label{fig:T0phasediagram}
\end{figure*}

{\sl Zero field phase diagram.} For orientation and to demonstrate the robustness of our methods we present in Fig.~\ref{fig:phasediagram} the zero-field phase diagram of the fully nested ($\phi=\pi/6$) model in the temperature ($T$)-interaction ($U$) plane obtained from single-site and cluster dynamical mean-field methods. Paramagnetic insulator, paramagnetic metal and antiferromagnetic insulator phases are found. The phase boundaries determined by the different methods are quantitatively similar almost everywhere, strongly suggesting that the results we find are insensitive to cluster effects. The only important difference is that, as is well known, the single-site DMFT method strongly overestimates the low $T$ critical $U$ needed to drive a paramagnetic metal-paramagnetic insulator phase transition; but it is important to note that the region of large difference occurs within the $120^\circ$-antiferromagnetic phase (i.e., below $T_N$) where the paramagnetic phase single-site DMFT calculation is irrelevant.

We remark that the calculations involve a mean-field approximation, so at finite $N_c$ the calculations do not capture the long-wavelength fluctuations that convert the transition to one of the Kosterlitz-Thouless type (for $\phi\neq0$) or push the transition temperature to zero (for the Heisenberg-symmetry $\phi=0$ case \cite{Mermin1966}). The mean-field temperature found here should be interpreted as setting the scale at which magnetic fluctuations become both strong and long ranged.

\begin{figure*}
		\centering
		\includegraphics{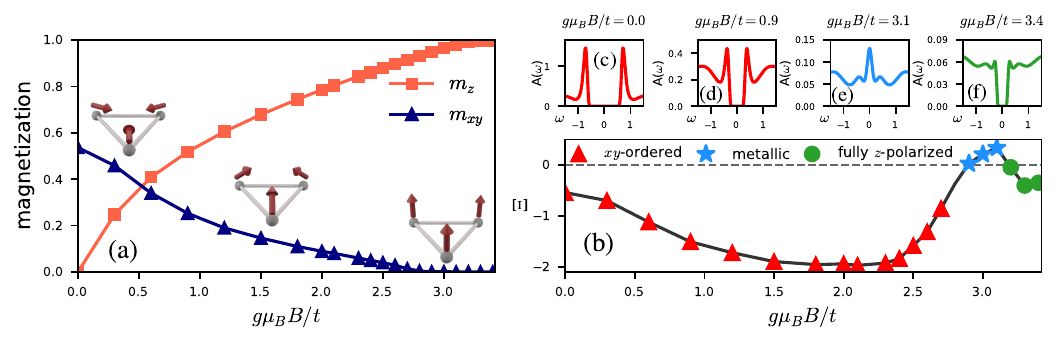}
		\caption{(a) Magnetic field (B) dependence of antiferromagnetic order parameter $m_{xy}$ (blue triangles) and spin polarization $m_z$ (orange squares). (b) shows $\Xi=-dA/dT$ which is positive (negative) in metallic (insulating) phases (see text). The local spectral functions (c)-(f), obtained with MaxEnt analytic continuation \cite{Kraberger2017}, confirm this classification. All quantities were computed at perfect nesting ($\phi=\pi/6$) for $U/t=4$ and $T/t\!=\!1/17$ with single-site DMFT.}
		\label{fig:observables}
\end{figure*}

{\sl Applied magnetic field at $T\!=\!0$.} Turning now to the effects of a magnetic field, in panel (a) of Fig.~\ref{fig:T0phasediagram} we show the phase diagram  in the magnetic field-interaction plane. This phase diagram was obtained at $T=0$ using the single-site VDAT, which has similar accuracy as the single-site DMFT, in the paramagnetic phase; however, spot checking the result with cluster methods and by allowing for magnetic order reveals that the single-site VDAT result for the full polarization line is quantitatively accurate.  A fully spin polarized, trivially insulating high-field phase is separated from a partially polarized phase by a transition line (sharp at $T=0$). At small-to-intermediate interaction the transition is continuous in the sense that the magnetization $m_z$ in the paramagnetic phase evolves smoothly up to the saturation value $m=1$, and our computed phase boundary agrees precisely with the Hartree-Fock result \cite{Zang2021} and results from exact diagonalization \cite{Wietek2022}. At larger interactions $U/t \gtrsim 6$ the transition changes to first order (meaning that $m_z$ jumps from a value less than $1$ to the saturated value) while the line deviates from the Hartree-Fock result and rolls over to the $\sim t^2/U$ saturation field expected for a Heisenberg magnet.

Also shown in the phase diagram is the paramagnetic (Mott) metal-insulator phase boundary. At the $\phi=\pi/6$ value used to construct Fig.~\ref{fig:T0phasediagram} the metallic phase is reentrant: in the small range $12.5\lesssim U/t\lesssim 14$ the zero $B$-field Mott insulator becomes metallic as the field is increased, before again becoming insulating. This reentrance is absent for  $\phi=0$   (see \cite{Supplemental} for the corresponding phase diagram) and it should be noted that cluster effects substantially change the single-site results for the Mott transition.

{\it Applied field at nonzero temperatures.} 
We now incorporate magnetic ordering and non-zero temperatures in the analysis. We focus on the interaction strength $U/t\!=\!4$, believed to be a suitable value for the description of the homobilayer WSe$_2$ \cite{Zang2022}. Panels (b)-(g) of Fig.~\ref{fig:T0phasediagram} show the Green functions and self-energies (minus their Hartree contributions) on the Matsubara axis, obtained from single-site DMFT at $T/t=1/17$. At low field strengths $g\mu_\text{B}B/t\!=\!1$, the system is insulating, indicated by the decrease in the imaginary part of the Green function at low Matsubara frequencies. This gap is opened by a strongly spin-dependent self-energy. At large field strengths $g\mu_\text{B}B/t\! \gtrsim \!3.4$ the system is fully polarized,  the lower $\sigma=\uparrow$ band is completely filled, and the combination of the magnetic field and the interaction opens a gap between the spin up and spin down bands. In between, e.g., at $g\mu_\text{B}B/t\!=\!3$ and temperature $T/t=1/17$, the $xy$-ordering is suppressed by the magnetic field, however, the system is not yet fully $z$-polarized, and, hence, the system is metallic.

We now turn to physical observables that can be obtained from these raw Green function data. Panel~(a) of Fig.~\ref{fig:observables} shows that as the Zeeman-field is increased from zero the staggered magnetization $m_{xy}$ decreases and the uniform  magnetization $m_z$ increases, indicative of the spin canting expected for a Heisenberg-symmetry magnet and sketched on the figure. At the value $g\mu_BB/t=2.8$ the staggered magnetization vanishes, but at this field the uniform magnetization $m_z<1$, indicating at this temperature a small window of paramagnetic partially polarized phase separating the antiferromagnet from the fully polarized state.

Panel~(b) of Fig.~\ref{fig:observables} examines the evolution of the electronic properties of these states, plotting $\Xi\!\coloneqq\!-dA/dT$, where $A \!=\!-\frac{1}{\pi T}G(\tau=\beta/2)$ is an estimate for the many-body density of states at the Fermi level. For the metallic state $\Xi\!>\!0$, whereas for the gapped  states $\Xi\!<\!0$. The magnetic state at different fields is shown as colored symbols. Densities of states obtained by analytically continuing the Matsubara axis Green function are shown for several points in panels~(c)-(f), confirming the identification of the different phases. We see that at this value of $U$ insulating and magnetically ordered behavior are closely linked.

\begin{figure*}
		\centering
		\includegraphics{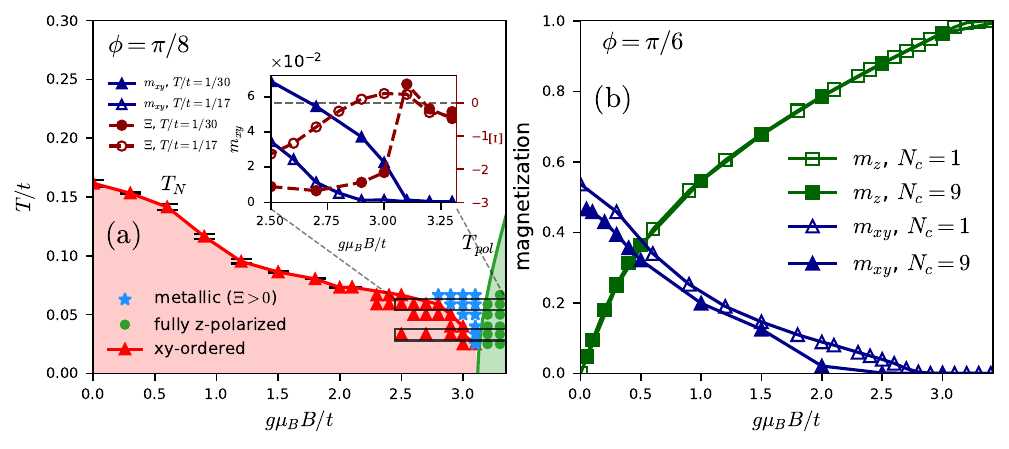}
		\caption{(a) Phase diagram in the plane of temperature and Zeeman magnetic field, indicating magnetic and metallic phases 
  computed for fixed $U/t\!=\!4$ and $\phi\!=\!\pi/8$ (imperfect nesting), calculated by DMFT. (b) Comparison of DMFT and 9-site CDMFT for $T/t=1/17$ and $\phi\!=\!\pi/6$ (perfect nesting).}
		\label{fig:BTphasediagram}
\end{figure*}

{\it $T-B$ phase diagrams and generalizations.} Fig.~\ref{fig:BTphasediagram} presents the phase diagram in the field-temperature plane at $U/t=4$ for a different phase angle ($\phi=\pi/8$), for which the nesting is imperfect and the van Hove singularity is removed from the Fermi surface. The analogous phase diagram for perfect nesting can be found in the Supplemental Material \cite{Supplemental}. Although some quantitative features change compared to $\phi\!=\!\pi/6$ (like a smaller onset of the magnetic ordering temperature at zero $B$-field), the two phase diagrams are qualitatively very similar: both show two types of magnetic ordering as well as an intermediate metallic phase (see inset of the right panel for a zoom into this regime).

The ordering temperature $T_\text{N}$ (red triangles), denoting the second order phase transition from a paramagnetic metal to a $120^\circ$ ordered insulator, is reduced upon the application of the external field. Interestingly, as already pointed out before (and in contrast to $T=0$), at non-zero temperatures an intermediate metallic phase (blue stars) appears with partial $z$-polarization. At even larger fields, the system  enters the fully polarized regime, which is insulating (green circles). At non-zero temperatures the $z$-magnetization is never completely saturated, hence we distinguish the partially polarized state from a `fully polarized' one using the criterion $m_{z}(B, T_{\text{pol}}) = 0.997$, which defines the boundary curve $T_{\text{pol}}(B)$ shown in Fig.~\ref{fig:BTphasediagram}.

We have found that as the temperature is decreased, the range of $B$ over which an intermediate metallic regime is observed decreases; the available evidence implies that at $T=0$ the entire range from $B=0$ up to the saturation field is $xy$-ordered and insulating, consistent with previous Hartree-Fock \cite{Zang2021} and exact diagonalization results \cite{Wietek2022}. Finally, let us note that non-local (spatial) correlations, neglected by DMFT, do not change the picture drastically. This can be inferred from the comparison (and qualitative agreement) of DMFT with 9-site CDMFT calculations in panel (b) of Fig.~\ref{fig:BTphasediagram} and is discussed further in the Supplemental Material \cite{Supplemental}.

{\it Conclusions and outlook.} In this paper we have investigated the full Zeeman magnetic field dependence of the metal-insulator and magnetic field phase diagram of the moir{\'e} Hubbard model (two dimensional triangular lattice Hubbard model with $xy$-magnetic anisotropy and nontrivial hopping phase). Our focus on the Zeeman (spin) coupling is motivated by the large g-factor and still not moderate lattice parameters available in current moir{\'e} systems. Future work will address the orbital effects of the field. Our results refine, extend and generalize the important early work of Laloux and Georges on the infinite dimensional Hubbard model \cite{Laloux94}. Our comparison of single-site and cluster dynamical mean-field approximations confirms that once magnetic ordering is allowed for, the single-site approximation provides a reasonably accurate solution even though the model is two-dimensional and that the VDAT method provides an extremely computationally efficient and highly accurate solution for ground state properties. Within this approximation we generally find, for both nested and non-nested  cases, that at $T=0$, the $B=0$ magnetic order and insulating behavior persists over the entire $B>0$ field range until the system becomes fully polarized, with the order parameter and transition temperature being gradually reduced by the spin canting. Metallic behavior is only found at nonzero temperatures for magnetic fields that suppress the magnetically ordered state to lower temperatures but are too weak to yield a fully spin-polarized state. Our results may be qualitatively compared to recent experiments on twisted WSe$_2$  which indicate that as the field is increased from zero the half filled material undergoes an insulator to metal transition at a field of about 2T, and becomes a fully polarized insulator at a field of $\approx 30$T \cite{Ghiotto23}. A precise comparison is difficult because the bandwidth cannot be unambiguously determined, but estimates from band theory \cite{Wang2020} and quantum oscillation measurements on a sample with a Moire lattice constant of $6.3$nm suggest at low hole density a mass of $\approx 0.4$m$_\text{e}$ implying $t\approx 2$meV. Use of the band theory $g$-factor $\approx 6$ and $U/t=4$ would then lead to a fully polarized state at about $30T$, consistent with experiment. However, the wide field range yielding metallic behavior is inconsistent with the theory. Whether this calls for a multi-orbital or non-local interaction extension of the MHM (analogously to an extended Hubbard model, see, e.g., \cite{Gneist2022,Zhou2022,Mazza2022}) is to be clarified by future studies, as well as the role of superconductivity \cite{Belanger2022,Klebl2023,Wu2023} and its connection to experiments.

{\it Acknowledgements.}
The authors are grateful for fruitful discussions with Sabine Andergassen, Laura Classen, Lorenzo Del Re, Antoine Georges, Henri Menke, Michael Scherer, and Nils Wentzell. T.S., M.K., and M.F. acknowledge the hospitality of the Center for Computational Quantum Physics at the Flatiron Institute. The authors acknowledge the computer support teams at CPHT {\'E}cole Polytechnique and at the Flatiron Institute. We thank the computing service facility of the MPI-FKF for their support.
We gratefully acknowledge use of the computational resources provided by the Max Planck Computing and Data Facility and by the IDCS mesocenter hosted at {\'E}cole Polytechnique.  J.~Z. acknowledges support from the NSF MRSEC program through the Center for Precision-Assembled Quantum Materials (PAQM) NSF-DMR-2011738 and A.J.M. was supported by Programmable Quantum Materials, an Energy Frontier Research Center funded by the U.S. Department of Energy (DOE), Office of Science, Basic Energy Sciences (BES), under award DE-SC0019443. The work by T.A.M. was supported by the U.S. Department of Energy, Office of Science, Basic Energy Sciences, Materials Sciences and Engineering Division. An award of computer time was provided by the INCITE program. This research also used resources of the Oak Ridge Leadership Computing Facility, which is a DOE Office of Science User Facility supported under Contract DE-AC05-00OR22725. The work by Z.C. and C.A.M. was supported by the Columbia Center for Computational Electrochemistry. The Flatiron Institute is a division of the Simons Foundation.

\bibliography{MHM_Main.bib}

\end{document}


\title{Magnetism and Metallicity in Moir{\'e} Transition Metal Dichalcogenides\\
{\it -- Supplemental Material --}}

\author{Patrick Tscheppe}
\author{Jiawei Zang}
\author{Marcel Klett}
\author{Seher Karakuzu}
\author{Armelle Celarier}
\author{Zhengqian Cheng}
\author{Chris A. Marianetti}
\author{Thomas A. Maier}
\author{Michel Ferrero}
\author{Andrew J. Millis}
\author{Thomas Sch{\"a}fer}
\maketitle
In this Supplemental Material we detail our calculation procedures and show additional data. In Sec.~\ref{sec:magnetism} we define different magnetizations and show how we calculated the magnetic ordering temperatures. Sec.~\ref{sec:dispersion} gives the non-interacting dispersion for different $\phi$s. Sec.~\ref{sec:mit} details the procedure of determining the interaction-driven metal-insulator crossover. Sec.~\ref{sec:polarized} defines the procedure for calculations with applied field. Sec.~\ref{sec:methods} gives details about the used algorithms, especially cluster layouts. Sec.~\ref{sec:plots} shows additional phase diagrams referred to in the main text.

\clearpage
\section{Magnetic order}
\label{sec:magnetism}
\begin{figure*}[b!]
		\includegraphics[width=\textwidth]{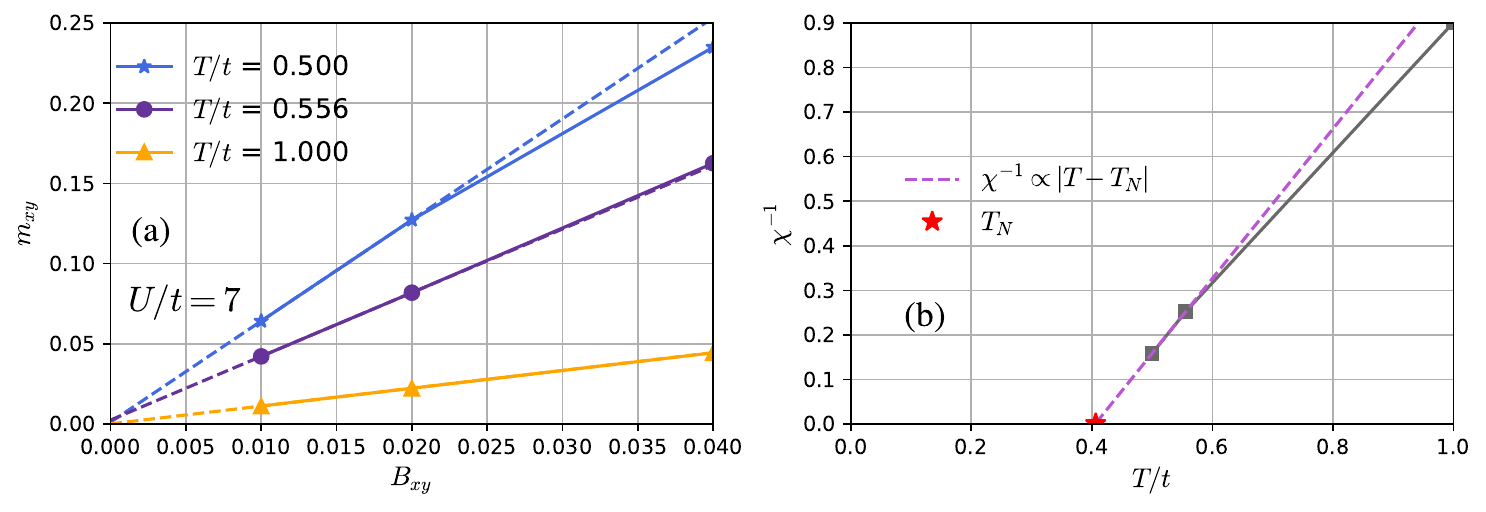}
        \captionsetup{justification=raggedright, singlelinecheck=false}
		\caption{Extraction of the N{\'e}el temperature $T_N$ for $xy$ ordering (we show data at $B=0$). (a) Staggered magnetization $m_{xy}$ as a function of applied in-plane field $B_{xy}$ for three different temperatures at $U/t = 7$ in 3-site CDMFT. From the slopes $\chi = \text{d}m_{xy}/\text{d}B_{xy}$ we extract the static susceptibility $\chi$. (b) Close to the transition the behavior of $\chi$ is well captured by a mean-field exponent $\chi^{-1} \propto \abs{T-T_N}$. By linear extrapolation we obtain the N{\'e}el temperature $T_N$.}
		\label{fig:chi}
\end{figure*}
At half-filling and for $\phi=\pi/6$ the low temperature magnetic ground state is characterized by a $120^\circ$ antiferromagnetic ordering pattern \cite{Pan2020, Zang2021, Zang2022}. As a consequence of broken $\mathrm{SU}(2)$ symmetry of the model the spins always align in the bilayer $xy$-plane and we define a corresponding staggered magnetization
\begin{equation}
    m_{xy} = e^{-i\vb{Q}\cdot \vb{R}_j} \qty(\expval{S_j^x} + i\expval{S_j^y}).
\end{equation}
Here $\vb{Q} = (- 4\pi/3a_\mathrm{M}, 0)$ and $a_\mathrm{M}$ denotes the moiré lattice spacing. Treatment of this order is greatly simplified by first applying a unitary transformation $c^\dagger_{j,\sigma} \rightarrow e^{i\sigma \vartheta_j/2}c^\dagger_{j,\sigma}$ mapping $S^x_j + i S^y_j \rightarrow e^{i\vartheta_j} (S^x_j + i S^y_j)$. We choose the phases $\vartheta_j$ such that $\vb{Q} = 0$ after the transformation, which is implemented by the replacement $\phi\rightarrow\phi + \pi/3$ \cite{Zang2021}. Competition between $xy$-order and strong $z$-polarization is then captured by the simpler order parameters $m_{xy}^2 = \expval{S_j^x}^2 + \expval{S_j^y}^2$ and $m_z = \expval{S_j^z}$. \par
At non-zero temperatures we use a response calculation to an externally applied field within the single-site DMFT and CDMFT approximations to obtain the ordering temperature $T_N$ for the $120^\circ$ state. The temperature dependence of the static susceptibility is well described by a mean-field exponent $\chi \propto \abs{T - T_N}^{-1}$ for all cluster sizes considered in this study. We therefore compute the staggered magnetization $m_{xy}$ at weak applied staggered fields $B_{xy}$ and at various temperatures and determine the susceptibility from the slope  $\chi = \text{d}m_{xy}/\text{d}B_{xy}$. The N{\'e}el temperature can then be extracted by linear extrapolation of $\chi^{-1}$. In Fig.~\ref{fig:chi} we show data for a 3-site CDMFT calculation to illustrate this procedure. 

\clearpage
\section{Dispersion}
\label{sec:dispersion}
\begin{figure}[b!]
    \centering
    \begin{subfigure}[b]{0.45\textwidth}
        \caption{$\phi = 0$ \linebreak$\sigma=\,\uparrow$\qquad\qquad\qquad\qquad\qquad\;\;\; $\sigma=\,\downarrow$}
        \includegraphics[width=\textwidth]{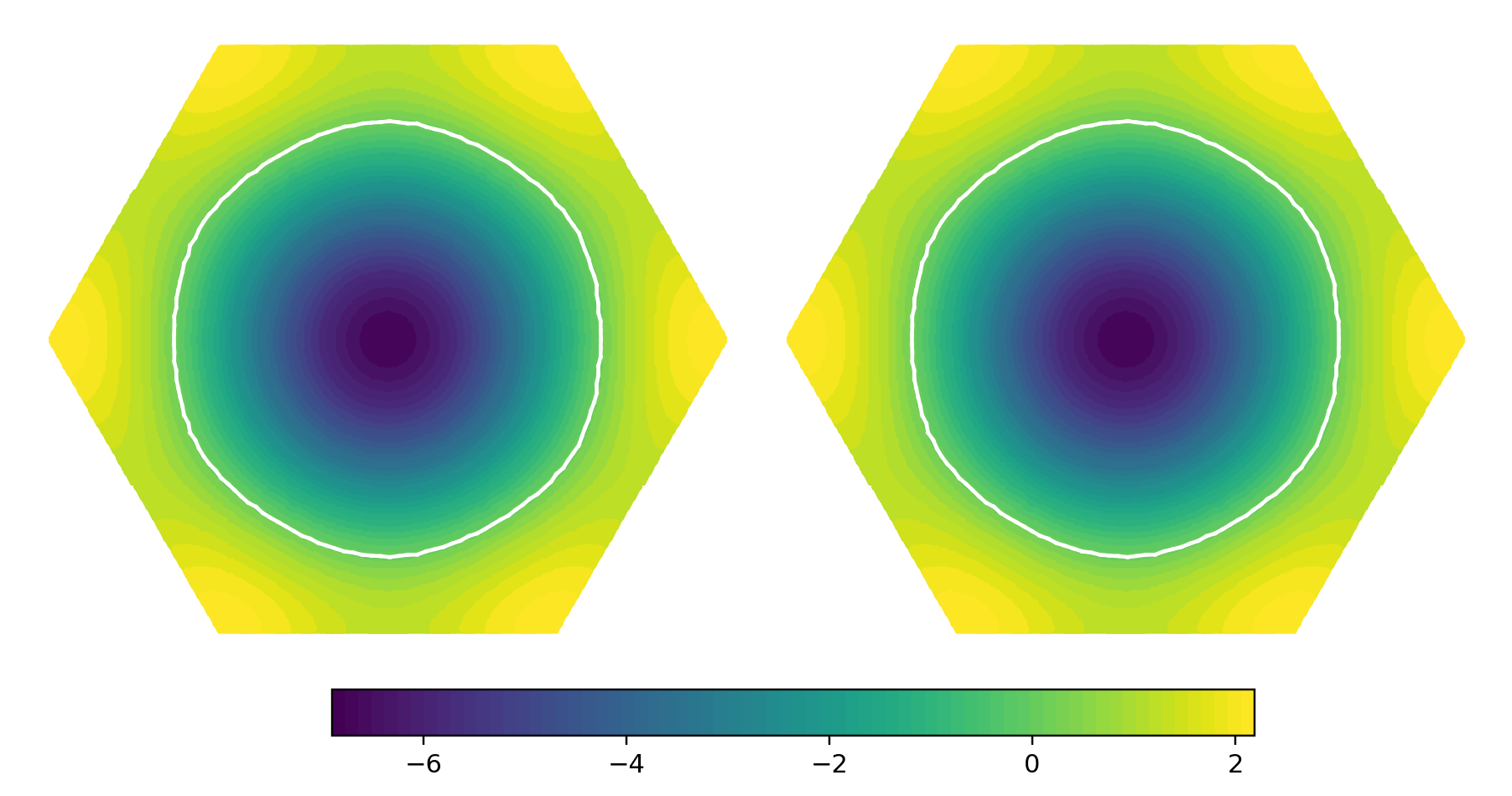}
        \vspace{-16pt}
        \label{fig:dispersion(a)}
        \caption*{\scalebox{0.8}{$\textstyle\qquad(\varepsilon_{\vb{k},\sigma} - \varepsilon_{\mathrm{F}})/t$}}
        \vspace{-5pt}
    \end{subfigure}
    \begin{subfigure}[b]{0.45\textwidth}
        \caption{$\phi = \pi/8$\linebreak$\sigma=\,\uparrow$ \qquad\qquad\qquad\qquad\qquad\;\;\; $\sigma=\,\downarrow$}
        \includegraphics[width=\textwidth]{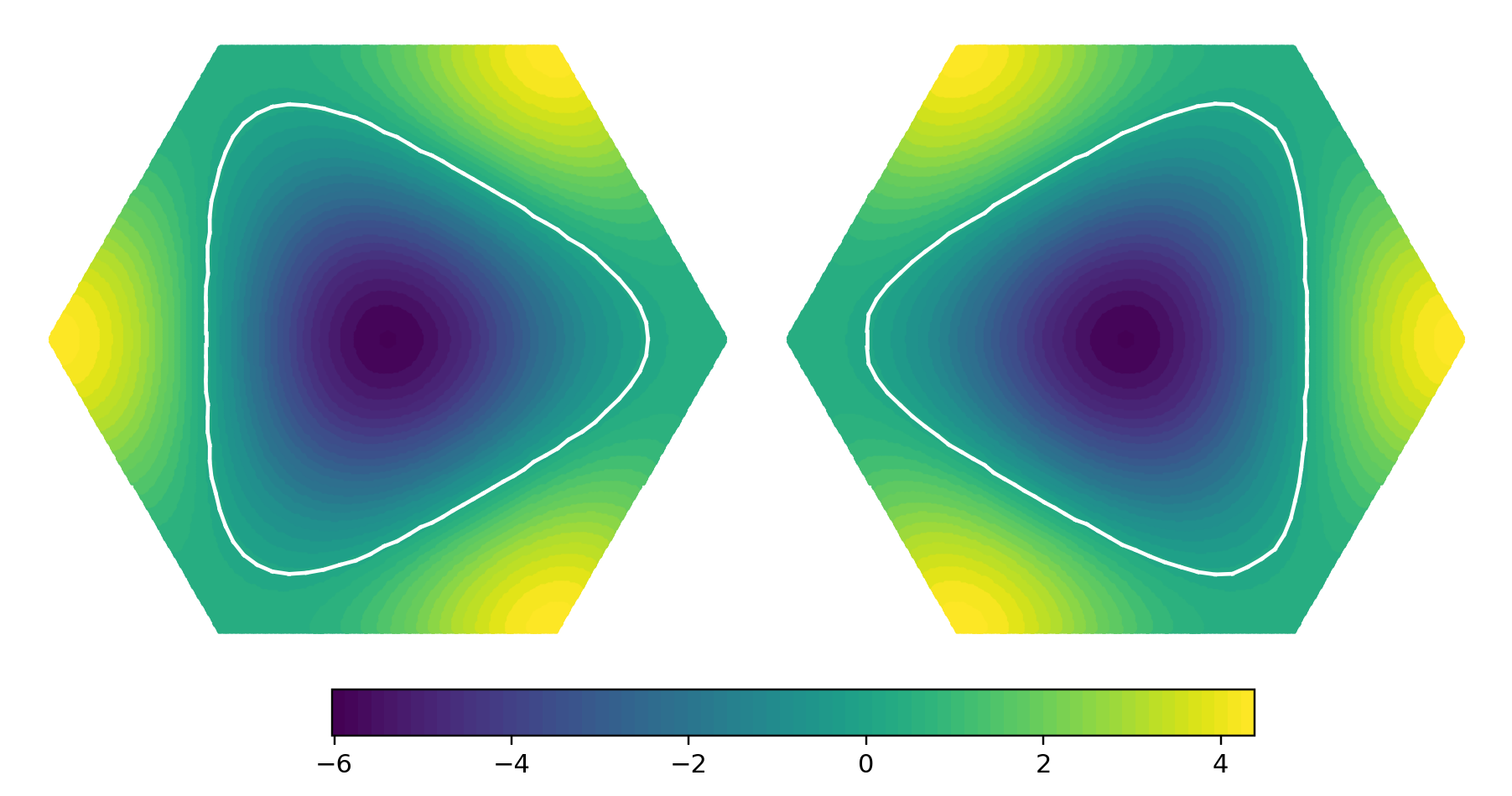}
        \vspace{-16pt}
        \label{fig:dispersion(b)}
        \caption*{\scalebox{0.8}{$\textstyle\qquad(\varepsilon_{\vb{k},\sigma} - \varepsilon_{\mathrm{F}})/t$}}
        \vspace{-5pt}
    \end{subfigure}
    \begin{subfigure}[b]{0.45\textwidth}
        \caption{$\phi = \pi/6$\linebreak$\sigma=\,\uparrow \qquad\qquad\qquad\qquad\qquad\;\;\; \sigma=\,\downarrow$}
        \includegraphics[width=\textwidth]{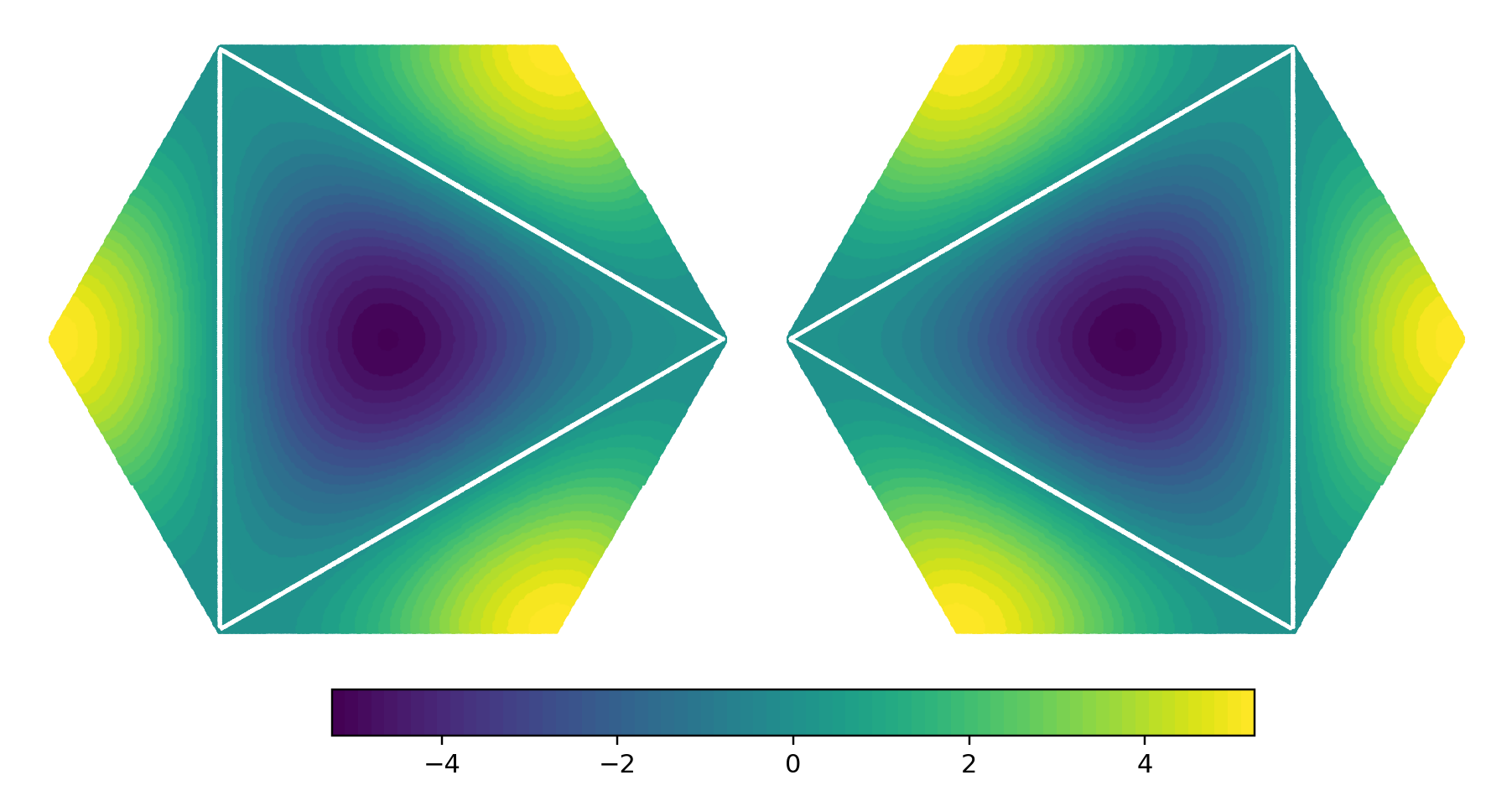}
        \vspace{-16pt}
        \label{fig:dispersion(c)}
        \caption*{\scalebox{0.8}{$\textstyle\qquad(\varepsilon_{\vb{k},\sigma} - \varepsilon_{\mathrm{F}})/t$}}
        \vspace{-5pt}
    \end{subfigure}
    \captionsetup{justification=raggedright, singlelinecheck=false}
    \caption{Non-interacting dispersion $\varepsilon_{\vb{k},\sigma} - \varepsilon_{\mathrm{F}}$ of the moiré Hubbard model for (a) $\phi=0$, (b) $\phi=\pi/8$ and (c) $\phi=\pi/6$ at half-filling. The Fermi surface $\varepsilon_{\vb{k},\sigma} = \varepsilon_{\mathrm{F}}$ is indicated in white.}
    \label{fig:dispersion}
\end{figure}
Fig.~\ref{fig:dispersion} shows the non-interacting dispersion $\varepsilon_{\vb{k},\sigma}$ of the MHM over the Brillouin zone for several values of $\phi$. At $\phi=0$ we recover the nearest-neighbor tight-binding dispersion on a triangular lattice. However, as $\phi$ is increased toward $\phi=\pi/6$ nesting develops in the spin up and down Fermi surfaces, which are now no longer identical and are related by time reversal symmetry. At $\phi=\pi/6$ there is both perfect nesting and a third-order van Hove singularity directly on the Fermi surface \cite{Zang2021, Zang2022}. For this reason the $120^\circ$ magnetic state persists down to infinitesimally small $U\to 0$. 

\clearpage
\section{Metal-insulator crossover}
\label{sec:mit}
\begin{figure*}[b!]
		\includegraphics[width=\textwidth]{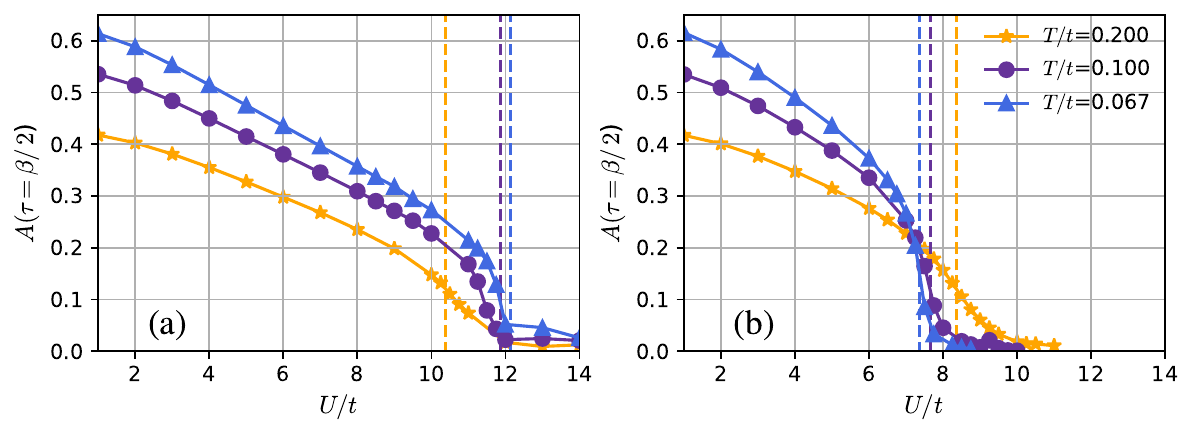}
        \captionsetup{justification=raggedright, singlelinecheck=false}
		\caption{Metal-insulator crossover at non-zero temperature in the perfectly nested moir{\'e} Hubbard model from single-site DMFT and 3-site CDMFT. (a) Zero-frequency spectral weight in DMFT computed using Eq.~\eqref{eq:A_loc}. (b) On the triangular 3-site cluster Eq.~\eqref{eq:A_loc} is applied to the local Green function of an arbitrary site since they are all equivalent by symmetry. In both cases the position $U_{c2}$ of the inflection point is used as a criterion for the metal-insulator crossover at non-zero temperature (vertical dashed lines). We observe that the inclusion of non-local correlations reverses the temperature-dependence of $U_{c2}$.}
		\label{fig:mott}
\end{figure*}

For all temperatures above the magnetic dome (see Fig.~1 in the main text) there is no sharp transition between the metallic and insulating state but rather a smooth crossover. As a criterion for this crossover we use the inflection point in the temperature dependence of the local (spin-summed) spectral function at zero frequency
\begin{equation}
    A_\mathrm{loc}(\omega=0) = -\frac{1}{\pi}\sum_{\sigma =\uparrow/\downarrow} \Im G_{\mathrm{loc},\sigma}(\omega=0)
\end{equation}
as a function of the interaction strength $U$. This quantity is not directly accessible from our imaginary time data, but it can be approximated by
\begin{equation}\label{eq:A_loc}
    A_\mathrm{loc}(\omega=0)\approx-\frac{1}{\pi T}\, G_{\mathrm{loc}}(\tau=\beta/2),
\end{equation}
which we found to agree well with a direct Matsubara frequency extrapolation $i\omega_n \rightarrow i0^+$ of $G_\mathrm{loc}(i\omega_n)$. In Fig.~\ref{fig:mott} we show this quantity both for the single site DMFT and 3-site CDMFT. Inside the magnetic dome the true solution of these methods is always insulating, which necessitates a restriction to the metastable paramagnetic solution in order to track the evolution of the crossover line to lower temperatures. In this study we compute $U_{c2}$ only, i.e. the critical interaction strength obtained by gradually increasing $U$ from small values within the metallic phase. At low temperatures the metal-insulator crossover can turn into a first-order transition, characterized by a hysteresis behavior of $A_\mathrm{loc}(\omega=0)$. We then expect to find a different value $U_{c1}$ for the transition when decreasing the interaction $U$. \\ While there is good quantitative agreement between all cluster sizes (including $N_c = 1$) as to the location of the high temperature crossover, larger differences become apparent at lower temperatures. In particular, as has been extensively studied in previous literature, the slope of the crossover line $U_{c2}(T)$ has a different sign in DMFT as compared to all clusters with $N_c > 1$ (see Fig.~1 of the main text).

\clearpage
\section{Metallicity in the polarized state}
\label{sec:polarized}
\begin{figure*}[b!]
		\includegraphics[width=\textwidth]{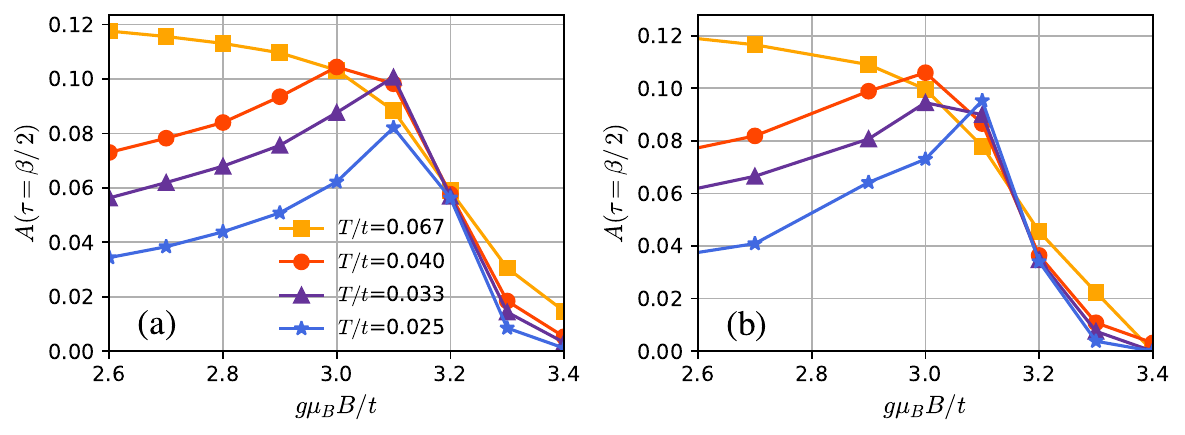}
        \captionsetup{justification=raggedright, singlelinecheck=false}
		\caption{Metallicity in single-site DMFT at $U/t=4$. (a) Local spectral weight at the Fermi energy for $\phi=\pi/6$ and four representative temperatures. For low and very large values of the applied Zeeman-field $B$ the spectral weight decreases monotonically with temperature. There is an intermediate metallic region around $g\mu_B B/t \approx 3.1$ where the spectral weight increases as the temperature is lowered. However, this region vanishes when $T/t < 0.033$. (b) Shows the same quantity at $\phi=\pi/8$, off of the perfect nesting condition. The qualitative behavior remains unchanged, while the metallic phase persists until lower temperatures.}
		\label{fig:spectral_weight}
\end{figure*}

\begin{figure*}[b!]
		\includegraphics[width=0.5\textwidth]{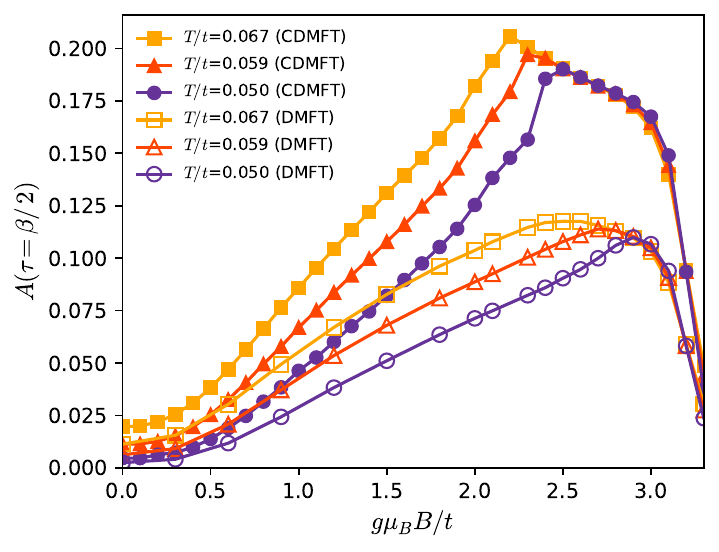}
        \captionsetup{justification=raggedright, singlelinecheck=false}
		\caption{Comparison of the local spectral weight at $\phi=\pi/6$ between single-site DMFT (filled symbols) and 9-site CDMFT (open symbols) as a function of the applied field and for different temperatures. While the slope $\dd{A}/\dd{B}$ changes almost smoothly at these temperatures in DMFT (at $g\mu_B B/t \approx 2.7$) we observe pronounced kinks at  $g\mu_B B/t \approx 2.3$ on the cluster.}
		\label{fig:A_cluster}
\end{figure*}

Once magnetic order is included,  at half filling the zero temperature ground state of the moir{\'e} Hubbard model is  insulating for $U$ greater than a critical value which vanishes for the perfectly nested case ($\phi=\pi/6$). However, for $U$ less than about $9t$, even if the $B=0$ ground state is insulating, at $T\neq 0$  we observe a reentrant insulator-metal-insulator transition as a function of the applied Zeeman-field at non-zero temperatures (Fig.~4 of the main text). As described in the main text we determine the metallic region by considering the temperature dependence of the Matsubara frequency estimator of the zero frequency spectral weight Eq.~\eqref{eq:A_loc}: if $A$ decreases to zero as $T$ is lowered we identify the phase as insulating; if it is temperature independent or tends to a nonzero value we identify the phase as metallic. In Fig.~\ref{fig:spectral_weight} this quantity is shown for applied Zeeman-fields $B$ close to the fully $z$-polarized region and at different temperatures. While the spectral weight decreases for small and very large values of $B$ when the temperature is lowered, we observe the opposite behavior in a small intermediate window at around $g\mu_\mathrm{B}B/t \approx 3.1$. Below a temperature of $T/t \approx 0.033$ this metallic phase is absent for $\phi=\pi/6$ in agreement with our zero temperature calculations which show no signs of metallicity. \\
Elaborating on the impact of non-local correlations on the insulator-metal transition, here we also compare the local spectral weights of single-site DMFT and 9-site CDMFT (Fig.~\ref{fig:A_cluster}). The transition between antiferromagnetic insulating and weakly temperature dependent metallic phase is concomitant with a change in the slope $\dd{A}/\dd{B}$ (at $g\mu_B B/t \approx 2.7$ and $\approx 2.3$ in DMFT and CDMFT, respectively) and these slope changes are much more pronounced in the cluster calculations (at $T/t = 0.059$ the slopes in the metallic regime are comparable in both cases, however right before the kink $\dd{A}/\dd{B} \approx 0.11$ in CDMFT v. the much smaller $\dd{A}/\dd{B} \approx 0.04$ in DMFT). \\
The currently available data indicate that the reentrant metallic phase is a temperature effect, with the phase remaining insulating at all $B$ for $T=0$; However, especially for the larger cluster calculations, further studies at lower temperatures (not technically feasible with currently computational resources) would be needed for providing a reliable extrapolation to the ground state.

\clearpage
\section{Numerical methods}
\label{sec:methods}
\begin{figure*}[b!]
		\includegraphics[width=0.8\textwidth]{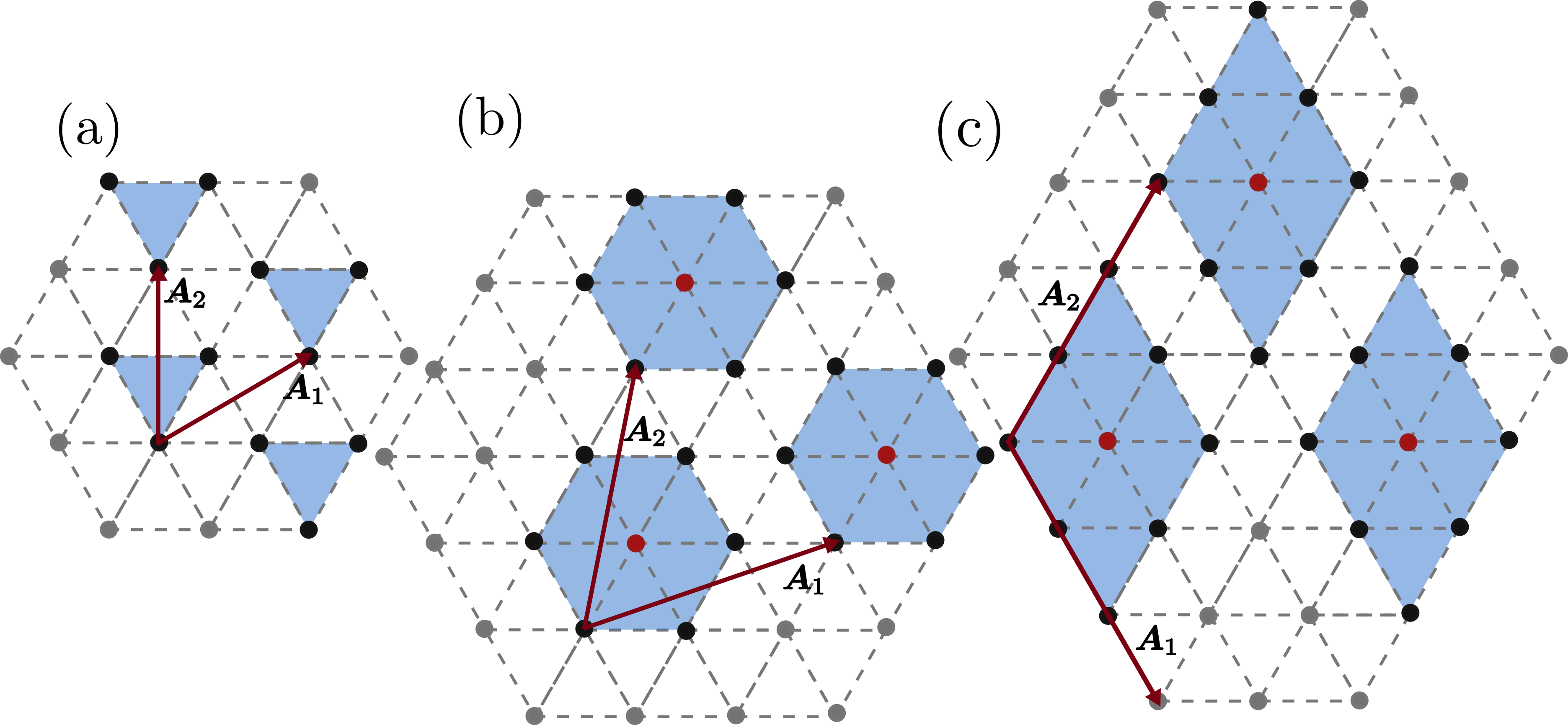}
        \captionsetup{justification=raggedright, singlelinecheck=false}
		\caption{Illustration of the various cluster geometries of size $N_c \in \{3, 7, 9\}$ used in the CDMFT calculations. (a) Due to lattice symmetry all sites in the three-site cluster are equivalent. (b), (c) In the seven and nine-site geometries there is a central site (red dots) whose nearest neighbors all belong to the cluster. By periodic repetition along superlattice vectors $\vb{A}_1, \vb{A}_2$ (dark red arrows) the clusters form a tiling of the triangular lattice.}
		\label{fig:clusters}
\end{figure*}

\subsection{Cellular dynamical mean-field theory (CDMFT)}
The single-site DMFT includes all local dynamical correlations between electrons. In order to systematically study the corrections due the to non-local correlations we here employ its cellular cluster extension CDMFT \cite{Kotliar2001, Maier2005} for three different cluster sizes having $N_c = 3,7$ and $9$ sites, respectively. In this approach the cluster degrees of freedom take the place of the DMFT single-site impurity, which is used to self-consistently approximate the lattice dynamics. We show the geometry of these clusters along with their corresponding superlattice vectors $\vb{A}_1, \vb{A}_2$ in Fig.~\ref{fig:clusters}. \\ Local quantities (such as magnetization and spectral weight) extracted from the center converge faster with cluster size \cite{Klett2020}. For this reason we show only data at the central site of the 7-site and 9-site clusters in the main text. \\
We solved the self-consistent cluster impurity model using the interaction expansion based continuous time quantum Monte Carlo solver \textsc{CT-INT} as implemented in the TRIQS framework \cite{TRIQS}.
\clearpage
\subsection{Dynamical cluster approximation (DCA)}
\begin{figure*}[b!]
		\includegraphics[width=0.4\textwidth]{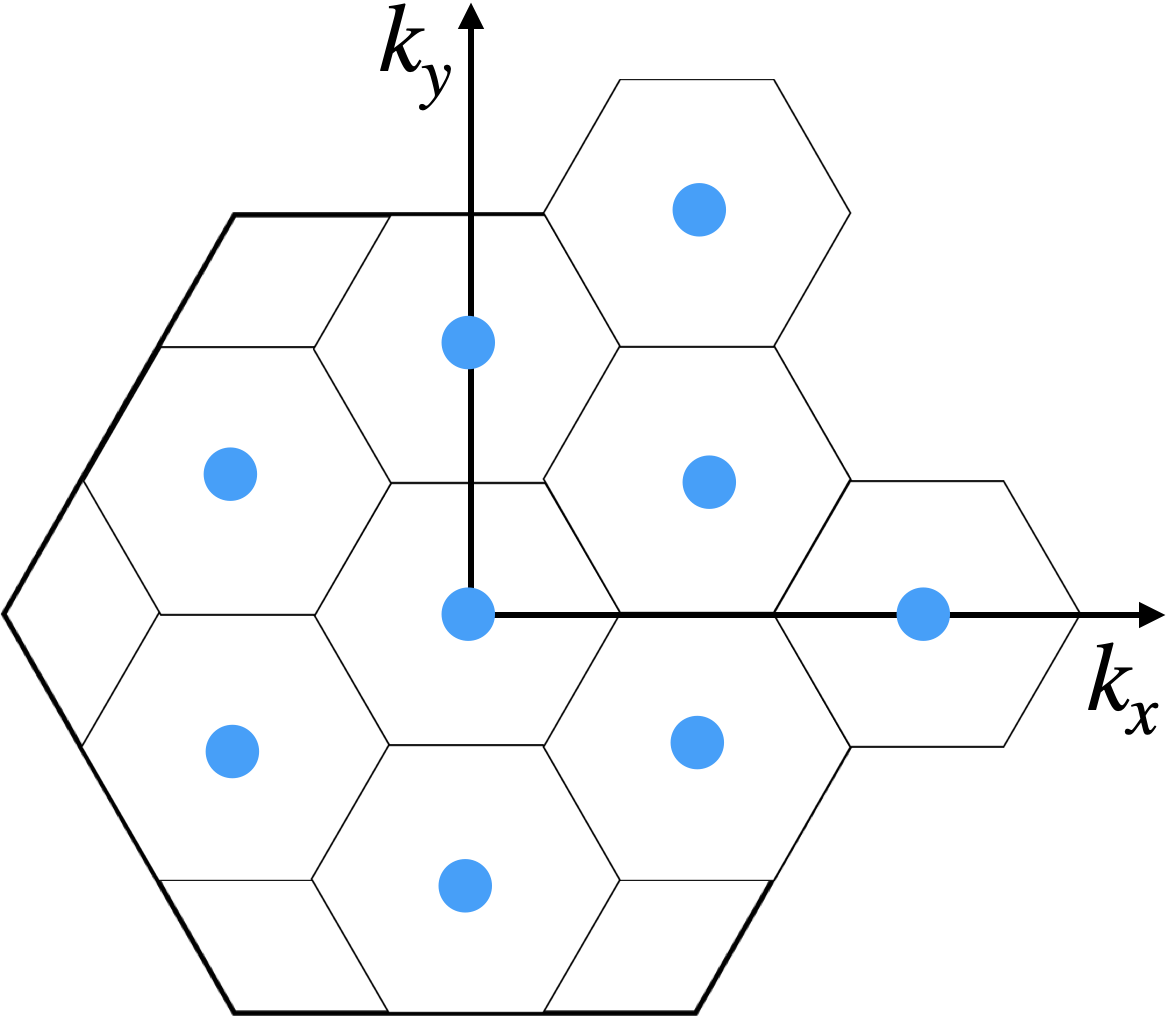}
        \captionsetup{justification=raggedright, singlelinecheck=false}
		\caption{Illustration of the triangular lattice Brillouin zone and the $N_c\!=\!3\times 3\!=\!9$ momentum space patches used in the DCA calculations.}
		\label{fig:DCAclusters}
\end{figure*}

DCA is an embedding method where an infinite size lattice is mapped onto a finite size cluster embedded in a dynamic mean-field, which is determined self-consistently~\cite{DCA_ref1}. In contrast to the CDMFT, this mapping is performed in momentum space by tiling the first Brillouin zone into $N_c$ patches as shown in Fig.~\ref{fig:DCAclusters}. This leads to an approximation in which short-range correlations within the cluster are treated accurately while longer-ranged correlations are approximated at the mean-field level. Here we use an $N_c\!=\!3\times 3\!=\!9$ cluster, for which the momentum space patches are shown in Fig.~\ref{fig:DCAclusters}.
To solve the effective cluster problem, we use a continuous-time auxiliary field quantum Monte Carlo (CT-AUX) solver ~\cite{CTAUX_ref1} as described in Ref.~\cite{DCA_code}. Unlike the CDMFT cluster, the DCA cluster is translationally invariant and the local single-particle Green function used to determine the critical $U_c$ shown in Fig.~1 in the main text is calculated as an average over the momentum space Green function. 

\subsection{Variational discrete action theory (VDAT)}
VDAT is a variational method to solve the ground state of a many-body system with an ansatz for the many-body density matrix, known as the sequential product density matrix (SPD), and a technique to evaluate the ansatz, known as the discrete action theory. The accuracy of the SPD is controlled by an integer $\mathcal{N}$. For the Hubbard model, the SPD with  $\mathcal{N}=1$ recovers the Hartree-Fock wave-function,  $\mathcal{N}=2$ recovers the Gutzwiller wave-function, and the large $\mathcal{N}$ limit recovers the exact ground state. In $d=\infty$, the SPD with any integer  $\mathcal{N}$ can be exactly evaluated within the discrete action theory via the self-consistent canonical discrete action theory (SCDA), which also provides a robust approximation for finite dimensions. The details are described in Refs.~\cite{Cheng2021,Cheng21,Cheng2022}. In this study, we used VDAT within the SCDA at $\mathcal{N}=3$, which is referred to as ``single-site VDAT" for brevity in the main manuscript. The code used to perform VDAT calculations is available \cite{VDATCode}.

\clearpage
\section{Additional plots}
\label{sec:plots}
\begin{figure*}[b!]
		\includegraphics[width=0.9\textwidth]{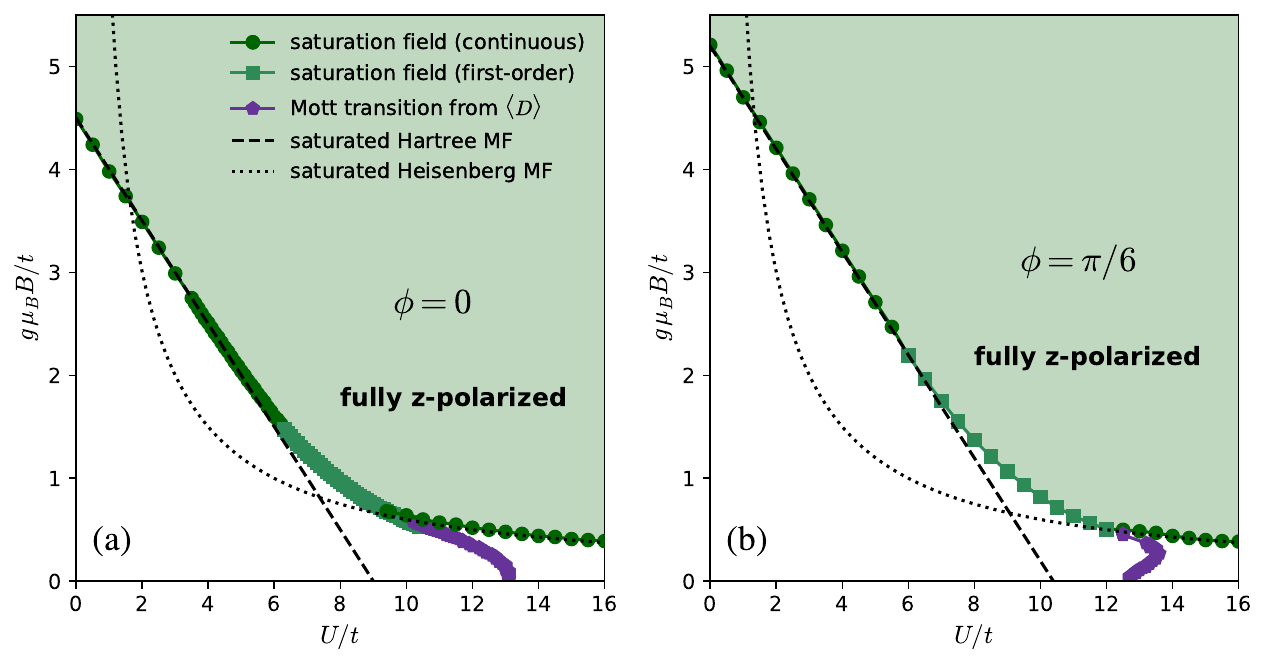}
        \captionsetup{justification=raggedright, singlelinecheck=false}
		\caption{Zero temperature phase diagram in the $U-B$-plane for the triangular lattice Hubbard model ($\phi=0$) and the perfectly nested moir{\'e} Hubbard model ($\phi=\pi/6$).}
		\label{fig:zero_temperature}
\end{figure*}

\begin{figure*}
		\includegraphics[width=0.9\textwidth]{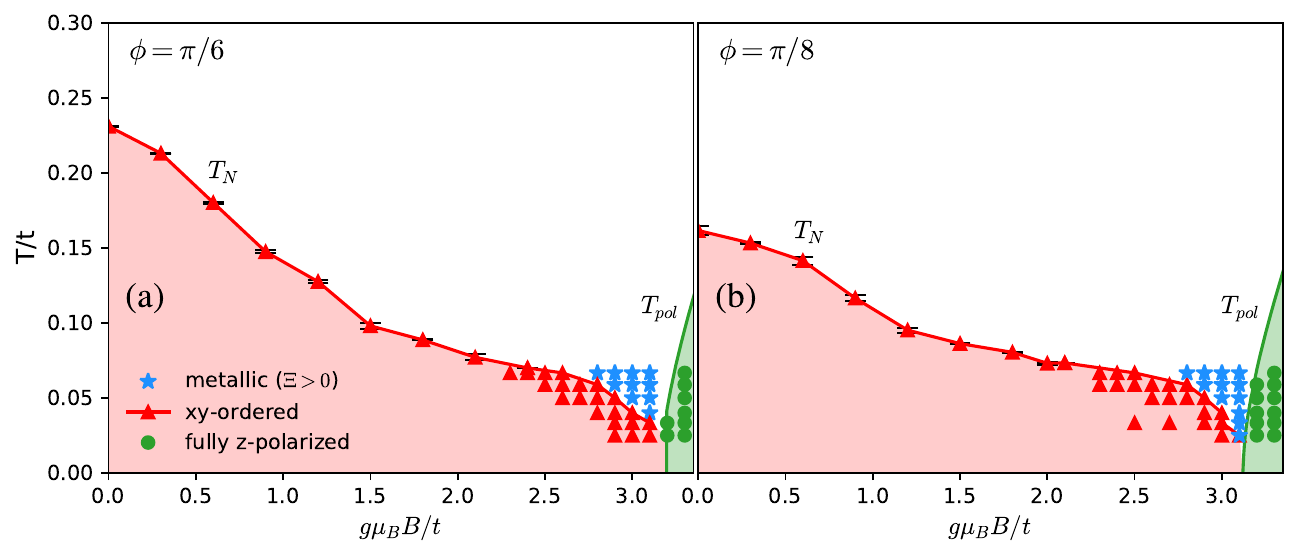}
        \captionsetup{justification=raggedright, singlelinecheck=false}
		\caption{Magnetic and metallic phases in the presence of an externally applied Zeeman-field $B$ for fixed $U/t\!=\!4$ and two different values of $\phi\!=\!\pi/6$ and $\phi\!=\!\pi/8$, calculated by DMFT.}
		\label{fig:BT_phi_pi_6}
\end{figure*}

In Fig.~\ref{fig:zero_temperature} we compare the zero temperature phase diagram of the perfectly nested moir{\'e} Hubbard model [$\phi=\pi/6$, panel (b)] with the spin-symmetric limit of the model [$\phi=0$, panel (a)]. Please note that, unlike in the perfect nesting situation discussed in the main text, the Mott transition as a function of the applied magnetic field is not reentrant at $\phi=0$. Still, in both cases the boundary of the fully $z$-polarized state follows the mean-field Heisenberg result $g\mu_\mathrm{B}B_{\mathrm{pol}} = 6t^2/U$ for $U$ beyond the Mott transition. \\
Fig.~\ref{fig:BT_phi_pi_6} shows the analogous plot to panel (a) of Fig.~4 of the main text also for $\phi=\pi/6$, i.e., the $B-T$ phase diagram at $U=4t$. Qualitatively the fully nested situation is very similar to the phase diagram at $\phi=\pi/8$. However, closer inspection reveals that the intermediate metallic phase is already completely suppressed at temperatures $T/t < 0.033 $, while it persists down to the lowest investigated temperature $T/t = 0.025$ for $\phi=\pi/8$.


\clearpage

\bibliography{MHM_Supplemental_Material.bib}